\documentclass[aps,preprint]{revtex4}
\usepackage{graphicx}
\usepackage{amsmath}
\usepackage{amssymb}
\usepackage{amsthm}
\usepackage{booktabs}
\usepackage{stmaryrd}
\usepackage{longtable}
\usepackage[figuresright]{rotating}
\usepackage{multirow}
\usepackage{dcolumn}
\usepackage{anysize}
\usepackage{color}
\marginsize{1.5cm}{1.5cm}{1.5cm}{1.5cm}
\usepackage{ulem}

\newcommand{\be}{\begin{equation}}\newcommand{\ee}{\end{equation}}
\newcommand{\bea}{\begin{eqnarray}}\newcommand{\eea}{\end{eqnarray}}
\newcommand{\bean}{\begin{eqnarray*}}\newcommand{\eean}{\end{eqnarray*}}

\begin{document}
\title{Crystal field and magnetism with Wannier functions: Rare-earth doped aluminum garnets.}
\author{E. Mih\'okov\'a\footnote{Fax:+420 2 312 3184, e-mail: mihokova@fzu.cz},  P. Nov\'{a}k, and V. V. Laguta\\~}
\affiliation{Institute of Physics, Acad.\ of Sciences of the Czech Rep., Cukrovarnick\'a~10, 162~53~Prague~6, Czech~Republic}
\date{\today}
\begin{abstract}
Using the recently developed method we calculate the crystal field parameters in yttrium and lutetium 
aluminum garnets doped with seven trivalent Kramers rare-earth ions. We then insert calculated parameters into 
the atomic-like Hamiltonian taking into account the electron-electron, 
spin-orbit and Zeeman interactions and determine the multiplet splitting by the crystal field as well as magnetic $\hat{g}$ tensors. 
We compare calculated results with available experimental data.

\end{abstract}

\maketitle

\section{Introduction}
To explain certain optical and magnetic properties of rare-earth ($R$) materials, determination of crystal field parameters (CFP) 
is essential. When sufficient experimental data are available CFP are usually determined by the least squares fit. Since the 
number of nonzero CFP depends on the site symmetry and may be as high as 27, such method often ends up being ambiguous. As 
a result, there has been a continuous effort to estimate CFP theoretically (for the review of various methods, see Ref.~\cite{survey1,survey2}). 

Recently, a novel theoretical approach to calculate CFP has been proposed~\cite{novak1}. An original motivation of 
the work was to explain the magnetic properties of rare-earth cobaltites $R$CoO$_3$ where available experimental data 
do not suffice to estimate CFP. 
The method starts with the density functional theory (DFT) based band 
structure calculation, followed by a transformation of the Bloch to Wannier basis. The local Hamiltonian is then expanded 
in terms of the spherical tensor operators. Resulting CFP are inserted in an atomic-like Hamiltonian involving the 
crystal field, 4$\!f-$4$\!f$ correlation, spin-orbit coupling and Zeeman interaction. The method does not suffer from 
the 4$\!f$ electron self-interaction (the difficult problem of DFT {\it ab initio} methods). The hybridization of the 
4$\!f$ states with other valence orbitals is taken into account {\it via} hybridization parameter $\Delta$, a single 
parameter of the method. In the recent paper \cite{novak4} a relatively simple way allowing to estimate this
parameter was suggested and applied to the $R$:LaF$_3$ system. 

The method has been extensively tested for rare-earth doped aluminates YAlO$_3$:R$^{3+}$ with orthorhombic perovskite 
structure~\cite{novak1}. Remarkable agreement of calculated and experimental data was achieved. Application to  gallates 
$R$GaO$_3$ and cobaltites $R$CoO$_3$~\cite{novak2} as well as manganites $R$MnO$_3$~\cite{novak3} followed. Even in 
these cases calculations provide a fair agreement with available experimental data.   

 Rare-earth doped aluminum garnets are widely used as laser materials and
scintillators due to which extensive experimental
data are available. In recent years magneto-optical properties of some
non-Kramers ions in garnet hosts have been studied.
High magneto-optical activity observed in some cases is of interest in
microwave amplifiers and generators. Magnetooptics of Tb$^{3+}$ and
Tm$^{3+}$ in Y$_3$Al$_5$O$_{12}$ (YAG) is studied
in~Refs. \onlinecite{GruberJAP103,ValievOM36} and~Ref. \onlinecite{ValievOS106}, respectively.
Magnetooptics of Eu$^{3+}$ in various garnets is reported in
~Refs. \onlinecite{ValievJRE29,ValievJRE31} and that of Pr$^{3+}$ in YAG
in~ Ref. \onlinecite{ValievJLum145}.

In this work we apply the new CFP method to rare-earth doped aluminum
garnets. After reviewing the theory and computational procedure, the next
section is devoted to analysis of the problem of determination of the
parameter $\Delta$ in more detail compared to~Ref. \onlinecite{novak4}. In the
following section we calculate crystals field parameters. We show examples
of multiplet splitting by the crystal field. We focus on Kramers ions and
calculation of their magnetic $g$ factors, yet more crystal-field sensitive
quantities, and compare them with experimental data.
\section{Theoretical approach and computational details} \label{s.theory}
The effective Hamiltonian describing 4$\!f$ states can be written as
\be
\hat{H}_{\textit {eff}}=\hat{H}_A+\hat{H}_Z+\hat{H}_{CF}
\,,
\label{ham}
\ee
where $\hat{H}_A$ is the spherically symmetric, free ion atomic-like Hamiltonian (for details see~Ref. \onlinecite{carnall}), 
while $\hat{H}_Z$ and $\hat{H}_{CF}$ are the  Zeeman interaction and crystal field Hamiltonian, respectively. 
In the Wybourne notation~\cite{wybourne} $\hat{H}_{CF}$ has the form
\be
\hat{H}_{CF}=\sum_{k=2,4,6}\,\sum_{q=-k}^{k}B_q^{(k)}\hat{C}_q^{(k)}
\,,
\ee
where $\hat{C}_q^{(k)}$ is a spherical tensor  operator of rank $k$ acting on the 4$\!f$ electrons of the $R$ ion. 
 The coefficients $B_q^{(k)}$ are the crystal field parameters.  
Hermiticity of $\hat{H}_{CF}$ requires  that $(B_{-q}^{(k)})^*=(-1)^qB_q^k$.

Calculation of crystal field parameters consists of four steps:
\begin{enumerate}
\item
Standard selfconsistent band calculation with 4$\!f$ states included in the core. The results yield the 
crystal field potential, subsequently used in the next step.
\item
The 4$\!f$ as well as oxygen 2$p$ and 2$s$ states are treated as the valence states in a nonselfconsistent 
calculation, all other states are moved away using the orbital shift operator. Relative position of 4$\!f$ 
and oxygen states is adjusted using the {\it hybridization} parameter $\Delta$ (a single parameter of the method).
\item
The 4$\!f$ band states are transformed to Wannier basis using the wien2wannier~\cite{wienwan} and wannier90~\cite{wan90} packages.
\item
Local 4$\!f$ Hamiltonian in the Wannier basis is extracted and expanded in a series of spherical tensor operators. 
The coefficients of expansion are the crystal field parameters.
\end{enumerate}

To perform the band structure calculations in steps 1 and 2 we used the WIEN2k package~\cite{wien} with 
implemented augmented plane waves + local orbital method. For the exchange correlation functional we applied 
the generalized-gradient approximation form~\cite{GGA}. We used experimental lattice parameters of 
Y$_3$Al$_5$O$_{12}$ (YAG) and Lu$_3$Al$_5$O$_{12}$ (LuAG)~\cite{lparam}, but 
the atomic positions within the unit cell were optimized for each $R$ substitution by minimizing the atomic forces. The unit cell in our calculations 
consisted of 80 atoms. The eigenvalue problem was solved in five points of the irreducible Brillouin zone 
and the number of basis functions was $\sim$ 7700 (corresponding to the parameter $RK_{\text max}$=6.5). 
The calculations were non-spin polarized. The atomic radii of R (Y,Lu), Al and O were 2.3, 1.7 and 1.55, respectively.

Once the crystal field parameters were determined we used the modified {\it lanthanide} package~\cite{lanth} 
to solve the eigenproblem for the Hamiltonian~(\ref{ham}). The results provide the multiplet splitting by the 
the crystal field. From the energy dependence on external magnetic field one extracts the $\hat{g}$ tensor 
(for more details see~Refs. \onlinecite{novak1,novak2}).

 \section{Hybridization parameter}
\label{sec:delta}
The parameter $\Delta$ appears due to hybridization between the rare-earth $4f$ states and the
valence states of its ligands. Our treatment of hybridization  is briefly described in Ref. \onlinecite{novak1}. 
In the $R$ containing orthorhombic perovskites \cite{novak1,novak2,novak3} remarkable agreement between
experimentally obtained spectroscopic data and the calculation was obtained by fixing the value of $\Delta$
at 0.6 Ry. With the same $\Delta$ also the magnetism was calculated and compared with experiment. However, less experimental data were
available and the agreement, though still satisfactory, was not as good as in case of spectroscopy.
In $R$:LaF$_3$ very good agreement between optical data and calculation was obtained for $\Delta$ = 0.4 Ry \cite{novak4} .
In the same paper $\Delta$ was estimated using a charge transfer energy
\begin{equation}
\label{eq:delta}
 \Delta \simeq E_{tot}(4f^{(n+1)},N_{val} - 1) - E_{tot}(4f^n,N_{val}),
\end{equation}
where $E_{tot}(4f^n,N_{val})$ is the total energy of the ground state of the system ($n_{4f}$ electrons in $4f$ shell of R ion and
$N_{val}$ electrons in the valence band), while $E_{tot}(4f^{(n+1)},N_{val} - 1)$ corresponds to the excited state in which
one of the valence electrons was transferred in the $4f$ electron shell. The hybridization parameter thus can be calculated
by performing two calculations with $4f$ electrons treated as the core states - the first one with $ 4f^n,N_{val}$, the second 
with $ 4f^{(n+1)},N_{val} - 1$ electron configurations. 

\begin{figure}
\includegraphics[width=12 cm]{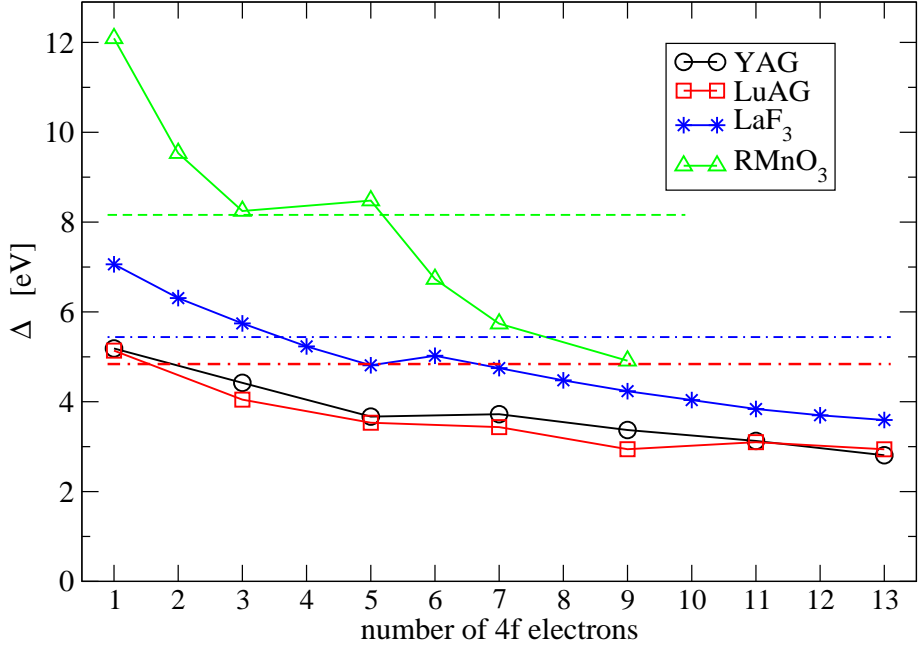}
\caption{Dependence of the hybridization parameter $\Delta$ on number of the $4f$ electrons calculated using eq. \ref{eq:delta}.
Dashed and dash-and-dotted lines correspond to the $\Delta$ values, which were adopted for calculations in $R$MnO$_3$ and $R$:LaF$_3$, respectively \cite{novak3,novak4}.}
\label{fig:Delta}
\end{figure}

Using the above equation we calculated $\Delta$ for the $R$ in question in both YAG and LuAG. The results, together with the
data for $R$:LaF$_3$ and orthorhombic perovskites $R$MnO$_3$, are shown in Fig.~\ref{fig:Delta}.

\begin{figure}
\includegraphics[width=12 cm]{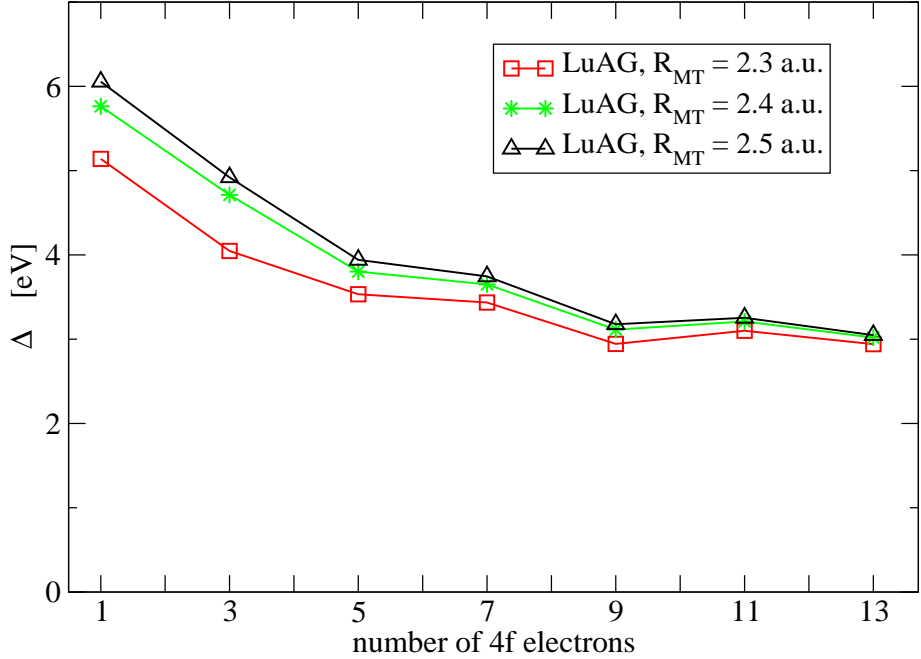}
\caption{$R$:LuAG. Hybridization parameter calculated for three values of the atomic sphere radius $R_{MT}$.}
\label{RMT}
\end{figure}


There are several problems connected with the above method. The first one is connected with the multiplet splitting of the $4f$ levels, 
which is not provided by the DFT calculation.
In principle this splitting may be obtained by the atomic-like program, we are using. However, there is a problem of double
counting of the electron-electron correlation, which would be difficult to overcome. We can only estimate that corresponding
uncertainty of $\Delta$ is on the order of the crystal field splitting i.e. less than $\sim$ 0.1 eV and it will change with $R$.  
Such scatter will become relatively more important in $R$:YAG and $R$:LuAG compared to
$R$:LaF$_3$ and $R$MnO$_3$, because $\Delta$ calculated from (\ref{eq:delta}) is smaller in garnets (cf. Fig.~\ref{fig:Delta}).

 The second problem is inherent to the open core calculations. Even
though the $4f$ electrons are well localized, small part of their density leaks out of the $R$ atomic sphere. Understandably,
this leakage is bigger for the $4f^{(n+1)}$ electron configuration. For $R$:LuAG we calculated $\Delta$ for three values
of the atomic sphere radius $R_{MT}$. The result is shown in Fig. \ref{RMT}.

Finally, note that eq. (\ref{eq:delta}) is based on the first order perturbation theory \cite{novak1}, thus higher order correction
is needed if the hybridization is important.
 
\section{Results\label{s.results}}
\subsection{CFP and multiplet splitting}
\begin{figure}
\includegraphics[width=9 cm]{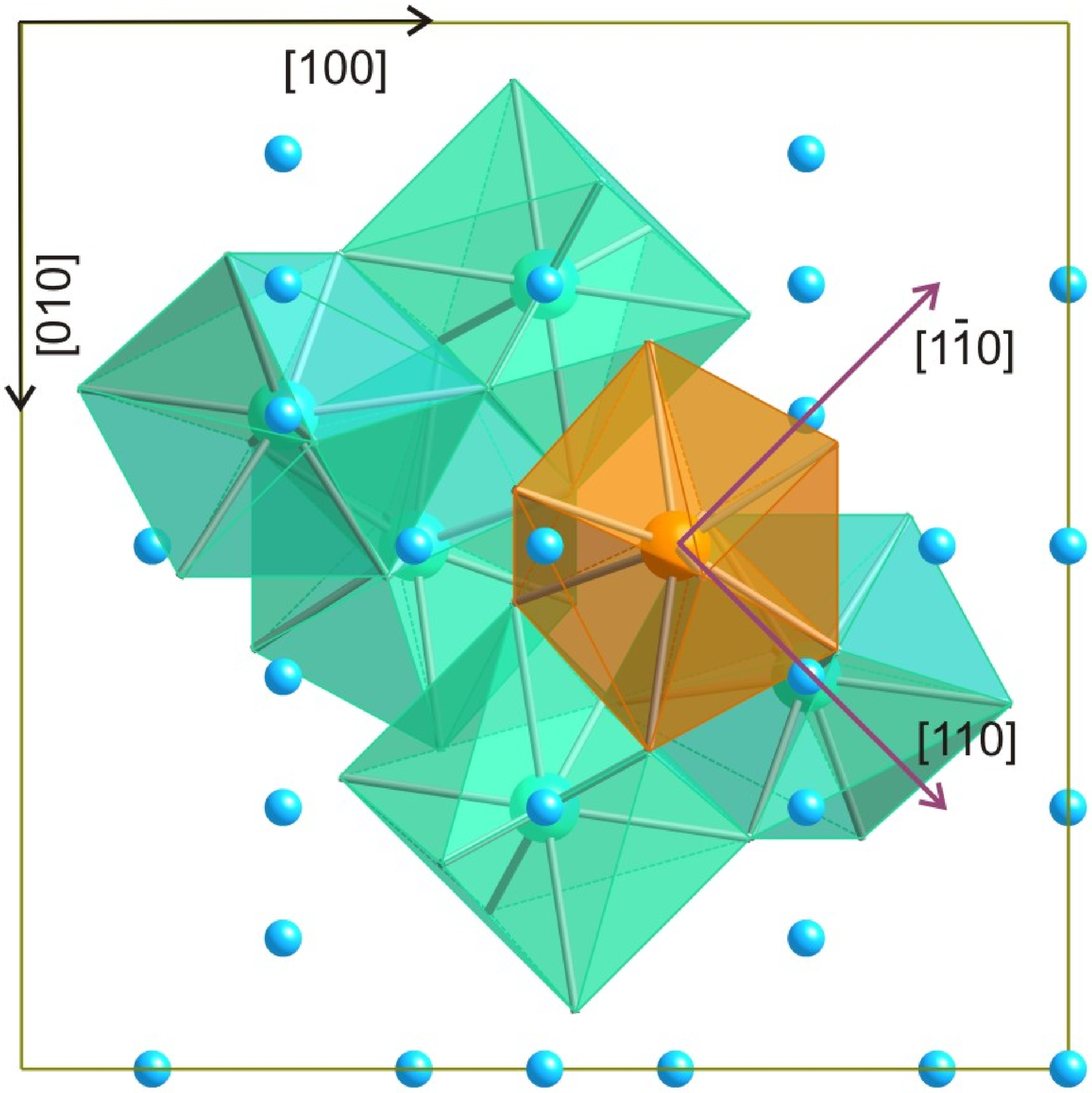}
\caption{ YAG structure with the highlighted dodecahedrally coordinated site.  Axes $a$ and $b$ are parallel 
to [1$\bar{1}$0] and [110] directions, axis $c$ is parallel to [001] direction.}

\label{fig:nn}
\end{figure}

Yttrium (Lutetium) aluminum garnets have a cubic structure belonging to the $Ia3d$ space group.  The unit 
cell contains eight molecular units 
Y$_3$Al$_5$O$_{12}$ (YAG) or Lu$_3$Al$_5$O$_{12}$(LuAG). The $R$ impurities are located on dodecahedrally coordinated Y (Lu) sites of 
$D_2$ symmetry, and nine real parameters are necessary to characterize the crystal field. For the dodecahedral sites, there are six 
possible coordinate system orientations, which result in six different 
(but spectroscopically indistinguishable) crystal field parametrizations. The coordinate system to which our CFP and $\hat{g}$ tensors
are referred is shown in Fig. \ref{fig:nn}.

\begin{figure}
\includegraphics[width=12 cm]{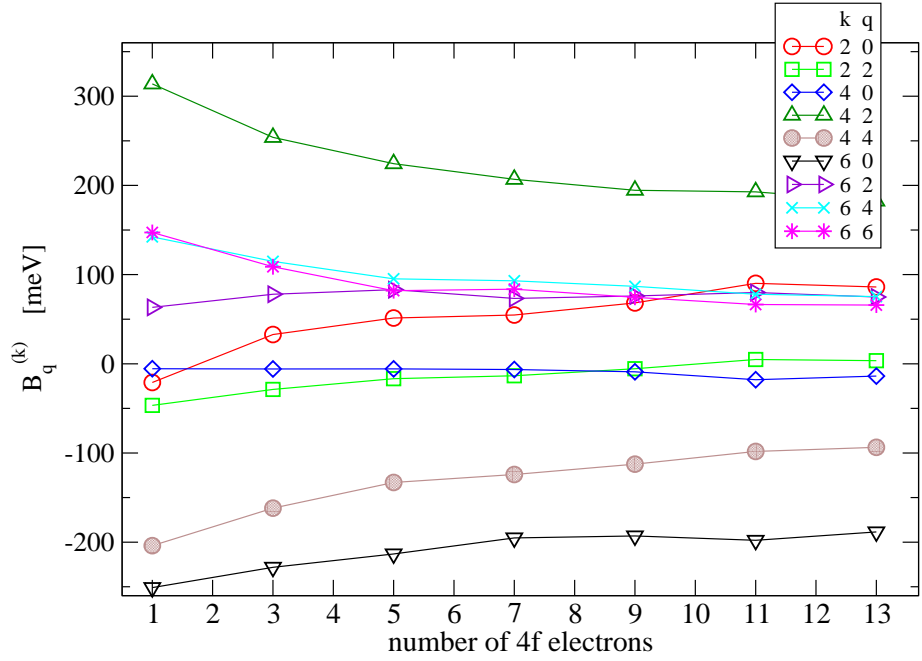}
\caption{YAG. CFP as a function of the number of $4f$ electrons for hybridization parameter $\Delta$=5.4 eV.}
\vskip 50 pt
\label{fig:cfp_nf}
\end{figure}

\begin{figure}
\includegraphics[width=12 cm]{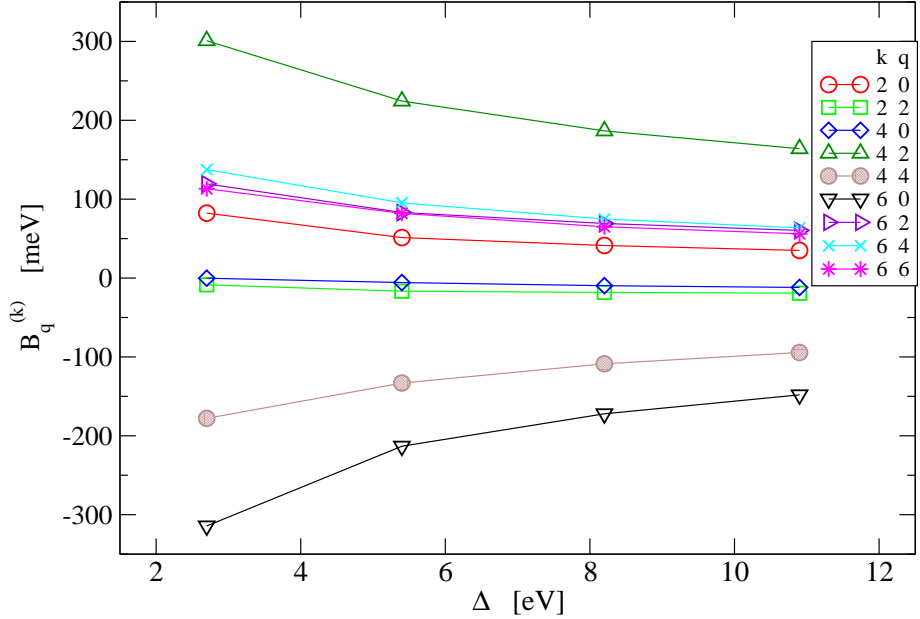}
\caption{Sm:YAG. CFP as a function of the hybridization parameter $\Delta$.}
\vskip 50 pt
\label{fig:cfp_D_sm}
\end{figure}

The magnitude of parameter $\Delta$ entering the calculation in the second step was varied between 2.7-10.9 eV (0.2-0.8 Ry)
with the step 1.36 eV (0.1 Ry).
As $\Delta$ decreases, the $4f$ levels get closer to the valence band, for $\Delta$ = 2.7 eV the calculation becomes less stable
for lighter $R$ and it crashes for Er and Yb. Similarly as in orthorhombic perovskites \cite{novak1,novak2,novak3} and LaF$_3$ \cite{novak4}
for fixed $\Delta$ the CFP change smoothly with the number of $4f$ electrons. In Fig. \ref{fig:cfp_nf} this
is documented for $\Delta$=5.4 eV and YAG. As a function of the hybridization parameter, CFP also change smoothly. 
An example for Sm:YAG is displayed in  Fig.~\ref{fig:cfp_D_sm}. The CFP for both YAG and LuAG and all seven $R$ are collected
in Table \ref{tab:cfp}. They were calculated taking $\Delta$=5.4 eV, which is the lowest value for which the
calculation runs smoothly for all $R$.

\begin{table}
\label{tab:cfp}
\caption{Crystal field parameters (in meV) in $R$:YAG and $R$:LuAG calculated for
hybridization parameter $\Delta$=5.4 eV.}
\begin{tabular}{rrrrrrrrrr}
\hline
 k  &q  &Ce:YAG  &Nd:YAG &Sm:YAG &Gd:YAG  &Dy:YAG   &Er:YAG   &Yb:YAG \\
 \hline
 2 &  0 &  -21.2 &   32.8 &   37.0 &   52.4 &   67.4 &  113.1 &   72.2 \\
 2 &  2 &  -46.6 &  -28.7 &  -23.8 &  -14.4 &   -5.6 &   15.4 &    2.1 \\
 4 &  0 &   -5.6 &   -5.8 &   -1.1 &   -5.8 &   -8.6 &  -22.9 &   -3.0 \\
 4 &  2 &  313.8 &  253.9 &  223.2 &  206.3 &  193.1 &  183.8 &  163.6 \\
 4 &  4 & -203.9 & -161.9 & -139.5 & -125.7 & -114.4 &  -87.3 &  -88.0 \\
 6 &  0 & -250.8 & -228.2 & -203.9 & -193.4 & -192.4 & -207.8 & -188.3 \\
 6 &  2 &   63.0 &   77.9 &   73.8 &   71.1 &   75.2 &   93.4 &   80.4 \\
 6 &  4 &  142.4 &  114.9 &  102.4 &   94.6 &   88.5 &   68.3 &   67.1 \\
 6 &  6 &  147.5 &  108.8 &   95.2 &   86.2 &   75.8 &   54.1 &   53.9 \\
\hline
   &  &Ce:LuAG  &Nd:LuAG &Sm:LuAG &Gd:LuAG  &Dy:LuAG   &Er:LuAG   &Yb:LuAG \\
 \hline
 2 &  0 &  -36.3 &   22.7 &   33.8 &   43.7 &   52.2 &   80.9 &   77.1 \\
 2 &  2 &  -33.0 &  -20.4 &  -14.2 &   -9.1 &   -3.5 &   18.8 &   17.8 \\
 4 &  0 &    4.6 &    1.9 &   -2.1 &   -2.0 &   -1.8 &  -28.9 &  -27.1 \\
 4 &  2 &  320.4 &  255.8 &  209.1 &  184.5 &  165.6 &  188.6 &  164.5 \\
 4 &  4 & -207.8 & -165.1 & -132.0 & -114.5 & -105.3 &  -97.1 &  -83.6 \\
 6 &  0 & -252.2 & -229.1 & -197.8 & -178.8 & -169.6 & -210.4 & -185.5 \\
 6 &  2 &   67.6 &   81.3 &   73.9 &   68.8 &   68.3 &   79.1 &   69.0 \\
 6 &  4 &  141.7 &  118.7 &   97.0 &   84.7 &   78.5 &   71.9 &   60.9 \\
 6 &  6 &  148.8 &  109.0 &   89.6 &   75.9 &   66.2 &   57.8 &   50.5 \\
 \hline
\end{tabular}
\end{table}

\begin{figure}
\includegraphics[width=12 cm]{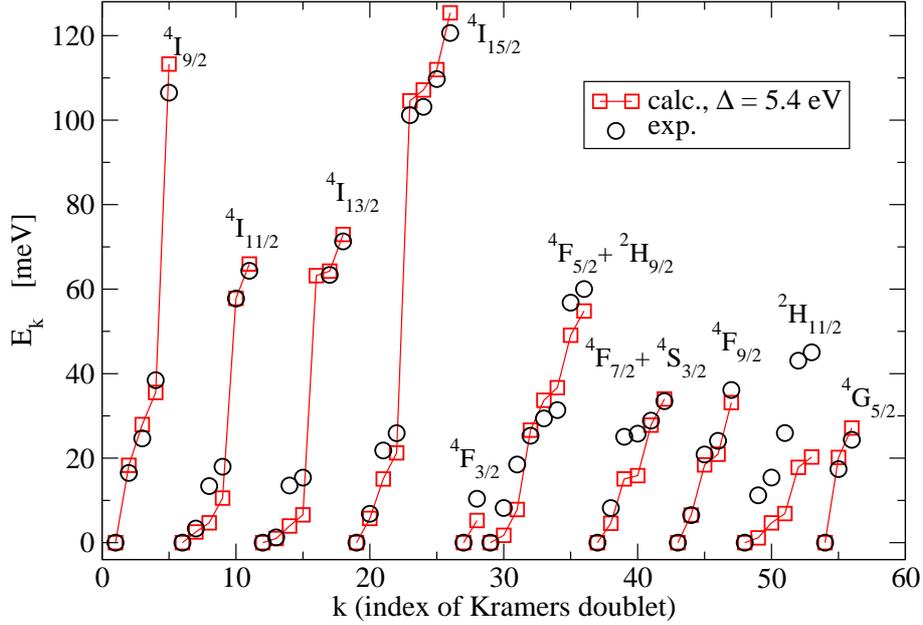}
\caption{Nd:YAG. Splitting of the lowest ten multiplets by the crystal field. The solid line is 
only to guide the eye. The lowest energy of the $j$-th multiplet ($j$=1,...,10) for calculated (experimental) results was reduced by 0(0), 255(248), 495(489), 
722(714), 1452(1418), 1567(1533), 1694(1656), 1851(1813), 1999(1952), 2020(2088) meV, respectively. Experimental data were taken from Ref.~\protect\onlinecite{burdickNd}.}
\label{fig:Nd}
\end{figure}

\begin{figure}
\vspace{1cm}
\includegraphics[width=12 cm]{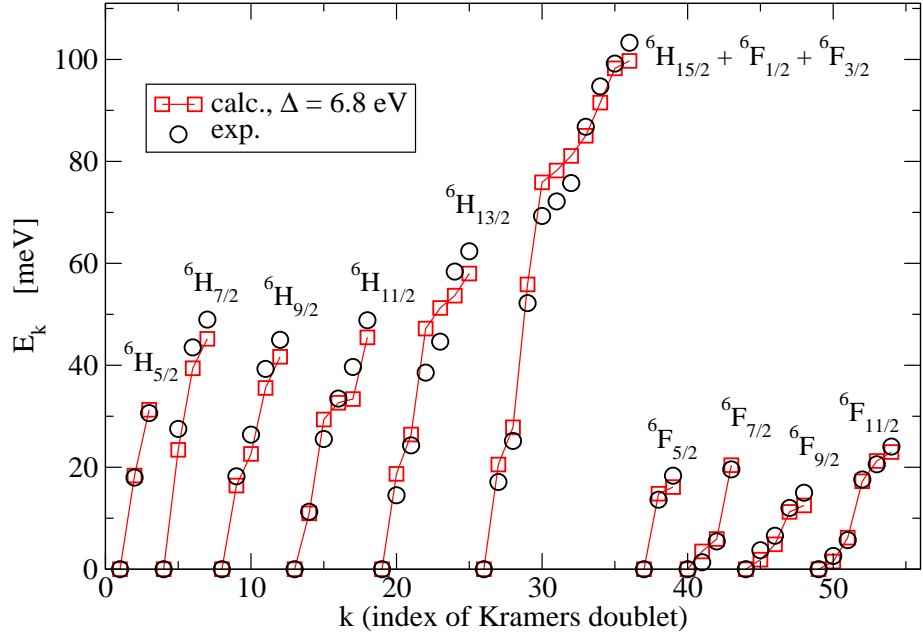}
\caption{Sm:YAG. Splitting of the lowest ten multiplets by the crystal field. The solid line is 
only to guide the eye. The lowest energy of the $j$-th multiplet ($j$=1,...,10) was reduced for calculated (experimental) results by 0(0), 
129(126), 284(279), 446(440), 609(603), 768(761), 909(813), 1017(831), 1164(833), 1331(848) meV, respectively. 
Experimental data were taken from Ref.~\protect\onlinecite{gruberSm}.}
\label{fig:Sm}
\end{figure}

\begin{figure}
\includegraphics[width=12 cm]{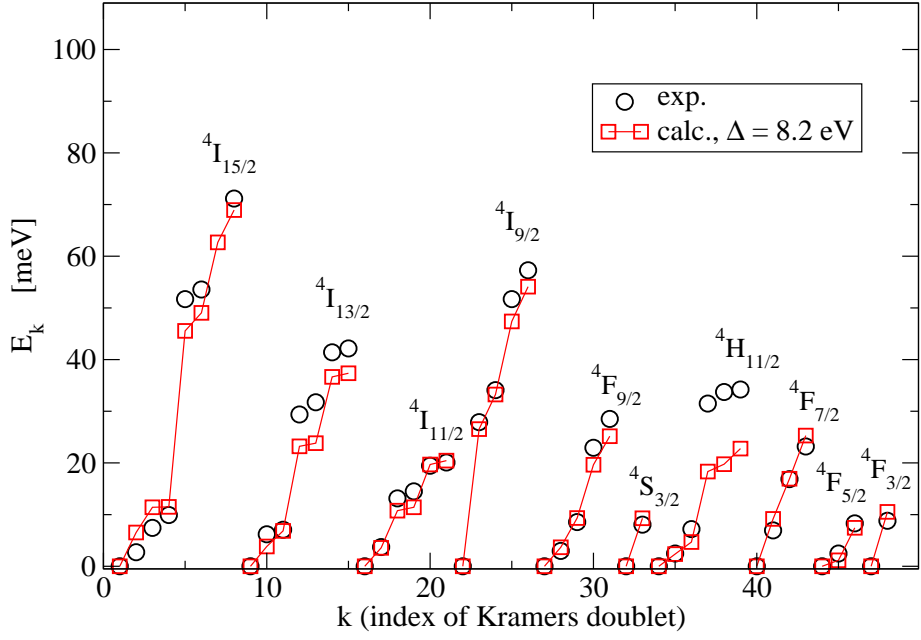}
\caption{Er:YAG. Splitting of the lowest ten multiplets by the crystal field. The solid line is 
only to guide the eye. The lowest energy of the $j$-th multiplet ($j$=1,...,10) for calculated (experimental) results was reduced by 0(0), 812(812), 1271(1272), 
1525(1525), 1895(1896), 2280(2281), 2367(2370), 2543(2543), 2755(2755), 2801(2803) meV, respectively. Experimental data were taken from Ref.~\protect\onlinecite{burdickEr}.}
\label{fig:Er}
\end{figure}

With the knowledge of CFP, modified program 'lanthanide' \cite{lanth} was used to calculate
the energy levels. Parameters of the free ion Hamiltonian were taken from Ref. \onlinecite{carnall}.
The agreement between theory and experiment is generally fairly good. Three examples, namely Nd$^{3+}$,  Sm$^{3+}$ and
 Er$^{3+}$ in YAG are shown in Figs. \ref{fig:Nd}, \ref{fig:Sm}, \ref{fig:Er}. For lighter $R$ the position of energy 
levels only slightly depends on the hybridization parameter, for heavier $R$ the dependence becomes stronger \cite{novak2}.

\subsection{$\hat{g}$ tensors}
 Axes $a, b, c$ of the system are principal axes of $\hat{g}$. In the experiment components $g_a, g_b, g_c$ are 
usually determined by analysis of the dependence of EPR spectra on the direction of external magnetic field. There are two sites
$R_1, R_2$ with the $c$ axis running along the [001] direction. Axes $a_1, b_1$, are parallel to $[1\bar{1}0]$ and [110], respectively. Axes
$a_2, b_2$ are obtained from $a_1, b_1$ by a $\pi$/2 rotation around $c$. This leads to an ambiguity in assessment of $g_a$, $g_b$ to
$R_1, R_2$ sites \cite{wolf}. On the other hand, in the calculation the assessment is unambiguous and it refers to the $R$ site
shown in Fig. \ref{fig:nn}.

To determine $\hat{g}$ the effective Hamiltonian (1) was diagonalized with gradually increasing external magnetic field $B$. Resulting
dependence of energies was then expanded up to the second power of $B$. The linear term provides $\hat{g}$. Comparing to energies in the 
zero-field, $\hat{g}$ is much more susceptible to parameters of the calculation, in particular to $\Delta$.
In Figs. 9-15 
the dependence $\hat{g}(\Delta)$ is shown for all $R$ in YAG and LuAG. As mentioned above,
for small $\Delta$ the calculations become unreliable, which results in fluctuations in $\hat{g}(\Delta)$ dependence. 
The calculation for $\Delta$ = 5.4 eV  and $R$:YAG, $R$:LuAG is compared with the experiment in Tables \ref{tab:gYAG} and \ref{tab:gLuAG}.

\begin{figure}
\includegraphics[width=12 cm]{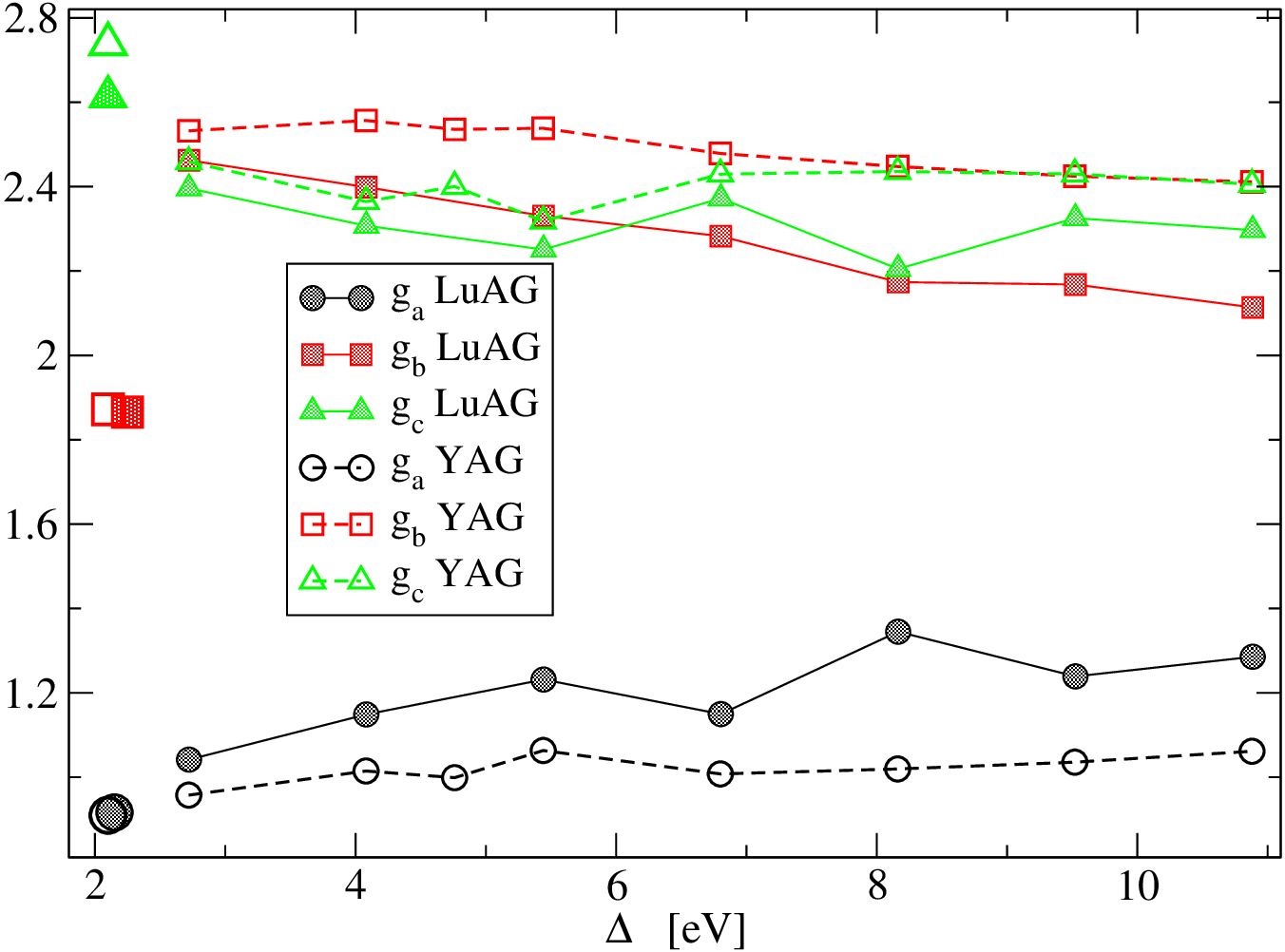}
\caption{Ce:YAG, LuAG. Principal components of $\hat{g}$ as a function of the hybridization parameter $\Delta$. 
Experimental data (enlarged symbols on the left side) are taken from Refs. \onlinecite{lewis,vedda}.}
\label{fig:ceg}
\end{figure}

\begin{figure}
\includegraphics[width=12 cm]{nd_g.eps}
\caption{Nd:YAG, LuAG. Principal components of $\hat{g}$ as a function of the hybridization parameter $\Delta$.
Experimental data (enlarged symbols on the left side) are taken from Refs. \onlinecite{nikolova,wolf}.}
\label{fig:ndg}
\end{figure}

\begin{figure}
\includegraphics[width=12 cm]{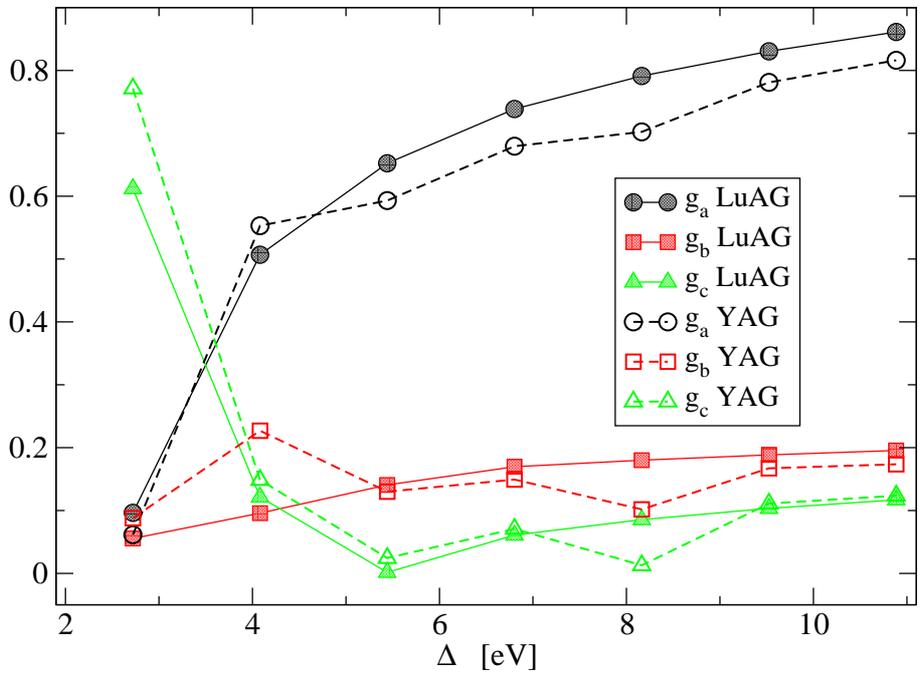}
\caption{Sm:YAG, LuAG. Principal components of $\hat{g}$ as a function of the hybridization parameter $\Delta$.
Experimental data are not available.}
\label{fig:smg}
\end{figure}

\begin{figure}
\includegraphics[width=12 cm]{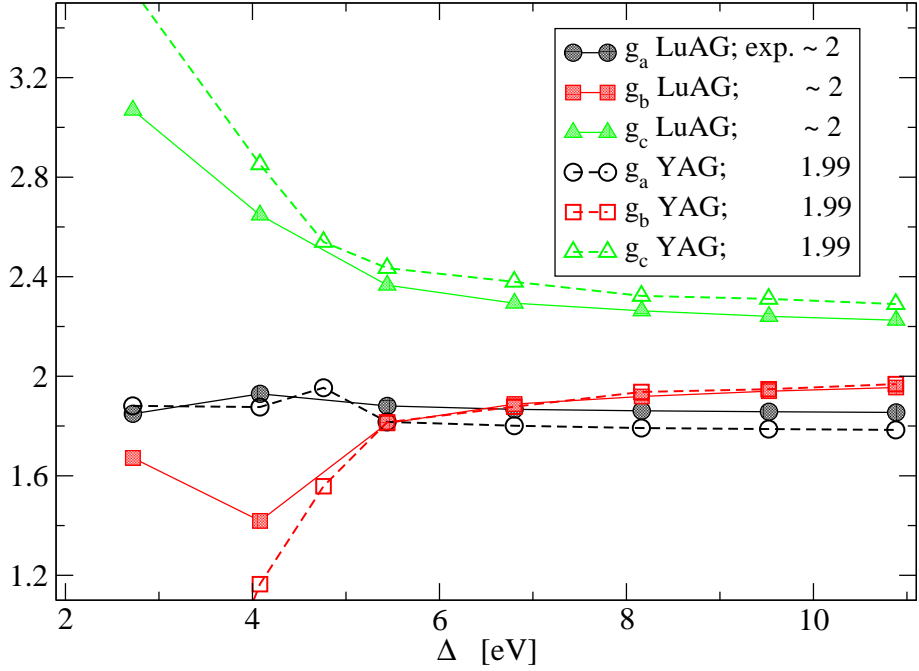}
\caption{Gd:YAG, LuAG. Principal components of $\hat{g}$ as a function of the hybridization parameter $\Delta$.}
\label{fig:gdg}
\end{figure}

\begin{figure}
\includegraphics[width=12 cm]{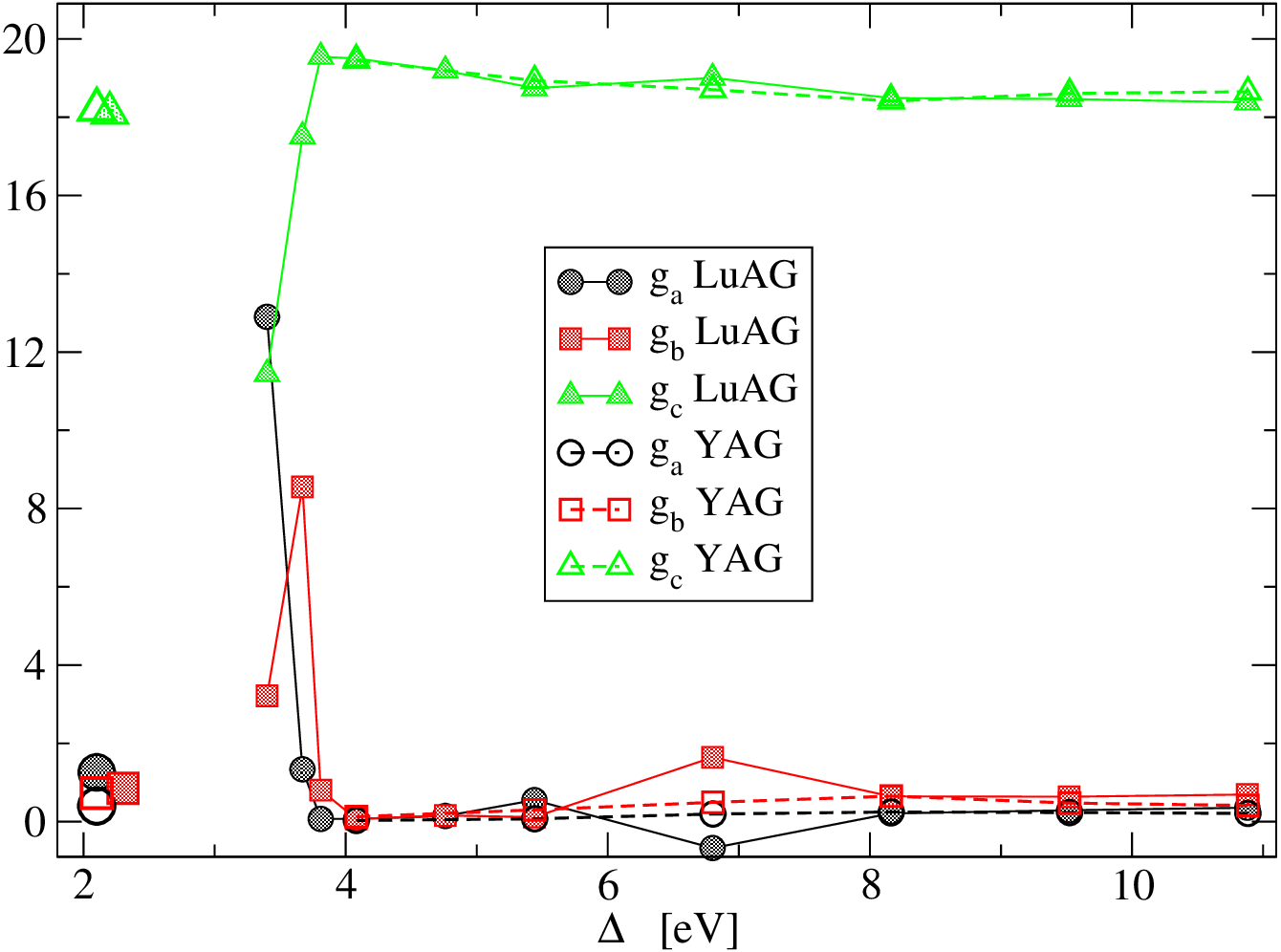}
\caption{Dy:YAG, LuAG. Principal components of $\hat{g}$ as a function of the hybridization parameter $\Delta$.
Experimental data (enlarged symbols on the left side) are taken from Ref. \onlinecite{CRC}.}
\label{fig:dyg}
\end{figure}

\begin{figure}
\includegraphics[width=12 cm]{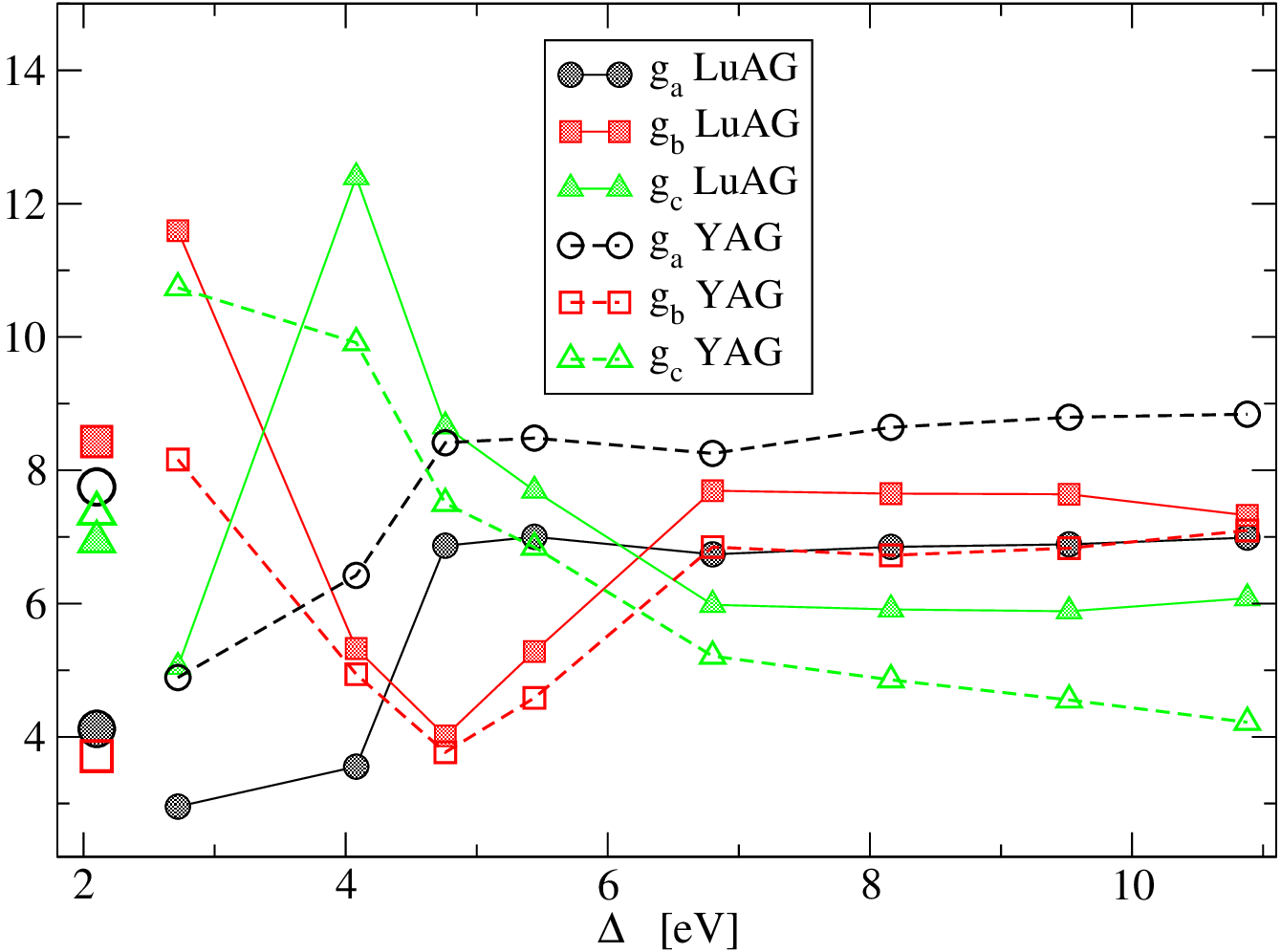}
\caption{Er:YAG, LuAG. Principal components of $\hat{g}$ as a function of the hybridization parameter $\Delta$.
Experimental data (enlarged symbols on the left side) are taken from Ref. \onlinecite{CRC}.}
\label{fig:erg}
\end{figure}

\begin{figure}
\includegraphics[width=12 cm]{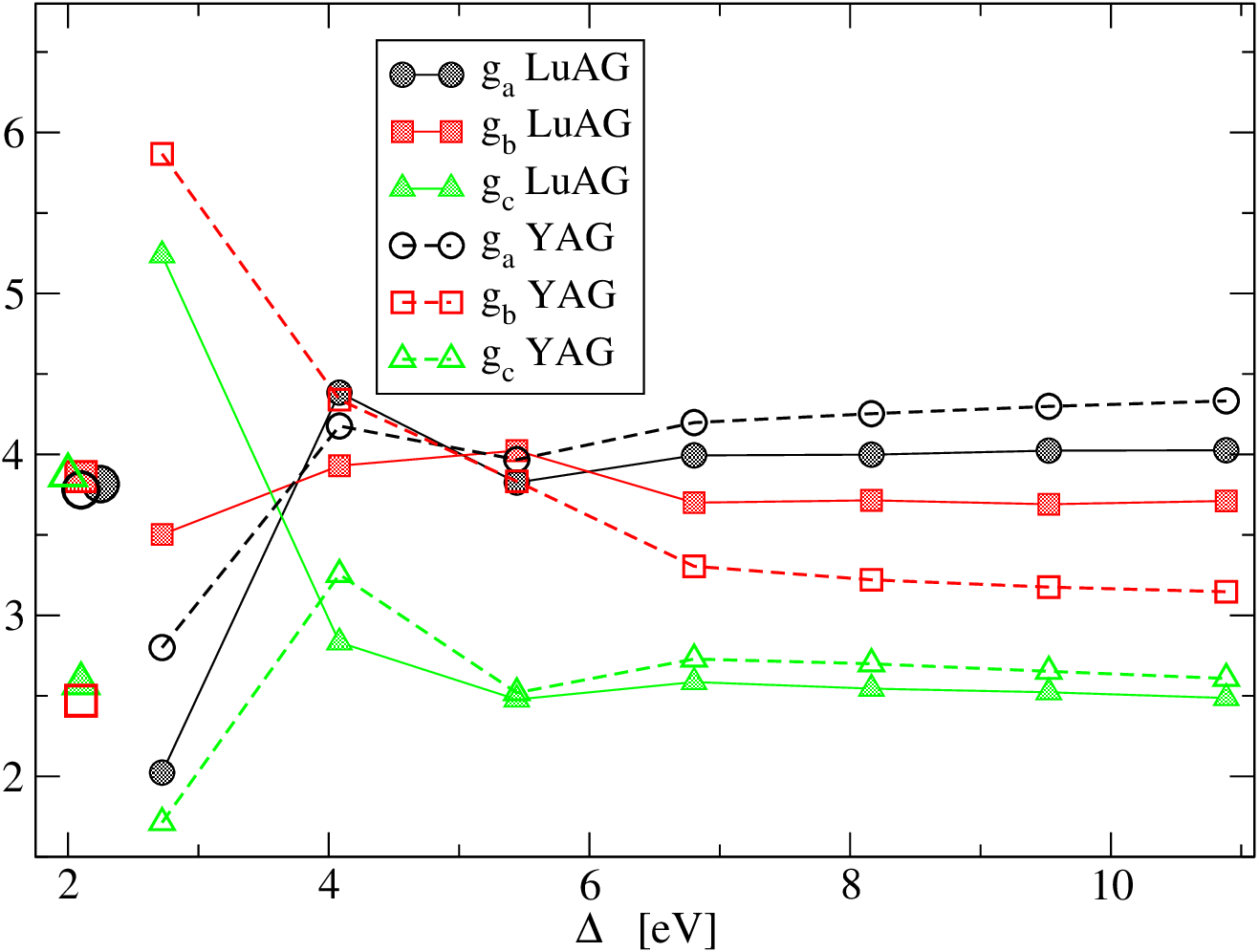}
\caption{Yb:YAG, LuAG. Principal components of $\hat{g}$ as a function of the hybridization parameter $\Delta$
Experimental data (enlarged symbols on the left side) are taken from Refs. \onlinecite{carson,laguta}.}
\label{fig:ybg}
\end{figure}


\begin{table}
\caption{$\hat{g}$ tensor components of the ground Kramers doublets along principal axes in 
YAG:R$^{3+}$ calculated for
hybridization parameter $\Delta$=5.4 eV, except for Er where  $\Delta$=4.8 eV. 
Experimental data refer to the work listed in the last column. Difference in per cent between calculated and experimental values is reported.}
\begin{tabular}{llllrlllrlllrl}
\hline
{\it R}  &\multicolumn{4}{c}{$g_a$}   &\multicolumn{4}{c}{$g_b$}   &\multicolumn{4}{c}{$g_c$}   &Ref.\\
     & calc. &exp.  &diff. & & calc. &exp.  &diff. & & calc. &exp.  &diff. &  &      \\
\hline
Ce  &  1.06 &  0.91 &16 & &  2.54 &  1.87 &36 & & 2.32 &  2.74 &15 & & [\onlinecite{lewis}] \\
Nd  &  2.16 &  1.74 &24 & &  1.74 &  1.16 &50 & & 3.52 &  3.91 &10 & & [\onlinecite{nikolova}] \\
Sm  &  0.59 &   --- &---& &  0.13 &  ---  &---& & 0.02 & ---   &---& & --- \\ 
Gd  &  1.82 &  1.99 &9  & &  1.82 &  1.99 &9  & & 2.43 &  1.99 &22 & & [\onlinecite{rimai}] \\
Dy  &  0.07 &  0.40 &83 & &  0.30 &  0.73 &59 & & 18.9 & 18.2  &4  & & [\onlinecite{CRC}] \\ 
Er  &  8.41 &  7.75 &9  & &  3.77 &  3.71 &2  & & 7.50 &  7.35 &2 & & [\onlinecite{CRC}] \\
Yb  &  3.97 &  3.78 &5  & &  3.83 &  3.87 &1  & & 2.52 &  2.47 &2 & & [\onlinecite{carson}] \\
\hline 
\end{tabular}
\label{tab:gYAG}
\end{table}

\begin{table}[h]
\caption{$\hat{g}$ tensor components of the ground Kramers doublets along principal axes in 
LuAG:R$^{3+}$ calculated for
hybridization parameter $\Delta$=5.4 eV, except for Er where  $\Delta$=4.8 eV. 
Experimental data refer to the work listed in the last column. Difference in per cent between calculated and experimental values is reported.}
\begin{tabular}{llllrlllrlllrl}
\hline
{\it R}  &\multicolumn{4}{c}{$g_a$}   &\multicolumn{4}{c}{$g_b$}   &\multicolumn{4}{c}{$g_c$}   &Ref.\\
     & calc. &exp. &diff.  &  & calc. &exp.  &diff.  &  & calc. &exp.  &diff  &  &    \\
\hline
Ce  &  1.23 &  0.92  &13 & &  2.33 &  1.87 &25  & &  2.25 &  2.61  &14 & & [\onlinecite{vedda}] \\
Nd  &  2.20 &  1.79  &23 & &  1.79 &  1.24 &44  & &  3.45 &  3.83  &10  & & [\onlinecite{wolf}] \\
Sm  &  0.65 &   ---  &---& &  0.14 &  ---  &--- & &  0.002&   ---  &---& &  --- \\ 
Gd  &  1.88 &   ---  &---& &  1.81 &  ---  &--- & &  2.37 &  ---   &---& & --- \\
Dy  &  0.55 &  2.29  &76 & &  0.11 &  0.91 &88  & & 18.74 &   16.6 &13 & & [\onlinecite{CRC}] \\ 
Er  &  6.87 &  4.12  &67 & &  4.01 &  8.43 &52  & &  8.66 &  6.93  &25  & & [\onlinecite{CRC}] \\
Yb  &  3.83 &  3.82  &0.3& &  4.02 &  3.72 &8   & &  2.48 &  2.57  &4  & & [\onlinecite{laguta}] \\
\hline 
\end{tabular}
\label{tab:gLuAG}
\end{table}

\section{Discussion}\label{sec:discussion}
As seen in Fig. 1, the hybridization parameter $\Delta$ for $R$ impurities in YAG and LuAG 
comes out smaller than in manganites and LaF$_3$. This represents a serious obstacle when trying
to get the best CFP, as for $\Delta$ smaller than $\simeq$ 4 eV the calculation does not always yield
reliable results. We traced the problem to the wannier90 package. Calculation with wannier90
provides the maximally localized Wannier functions, but it does not guarantee that they will be centered
on the crystal site of the $R$ impurity. Indeed, for $\Delta \simeq$ 4 eV the
functions are displaced for heavier $R$ and for still smaller $\Delta$ the displacement appears for all $R$.
As a consequence the symmetry is lost, all CFP are nonzero and for $ q \neq 0$ they are complex. A possible
remedy may be to use recently proposed scheme by Sakuma \cite{sakuma} ({\it symmetry adapted
Wannier functions}) or simpler, but less sophisticated method of {\it selectively localized Wannier functions}
\cite{wang}.

Despite the problem with $\Delta$ the agreement between calculated and experimental multiplet splitting is
very good as shown in Figs. \ref{fig:Nd},\ref{fig:Sm},\ref{fig:Er}. We found similar agreement also for other garnet systems.

In the past the CFP of the $R$ impurities in garnets were determined several times using either semiempirical methods
or the least squares fit to the optical data. Prominent groups that adopted this approach are those of Gruber and Burdick, which obtained rich
optical data and carefully analyzed them using the least squares fit. In Table \ref{tab:compar} we compare our CFP with those obtained by 
these groups for Nd:YAG \cite{burdickNd}, Sm:YAG \cite{gruberSm} and Er:YAG \cite{burdickEr}.

When considering $ B_{q}^{(k)}$ one should be aware of the fact that the spectroscopic methods do not
differentiate between the six crystallographically equivalent $R$ sites which, however, have different local
coordinate system. The local coordinate systems are connected by symmetry operations of the $Ia3d$ space group and 
to compare different sets of CFP corresponding symmetry operation has to be applied (see e.g. Refs. \onlinecite{morrison,burdickEr}). 

In Table \ref{tab:compar} are also compared the quantities $S_k$, which were introduced by Leavitt \cite{leavitt}
and which are  invariant with respect to the rotation of the coordinate system:
\begin{equation}
\label{eq:Sk}
   S_k = \left[\frac{1}{2k+1}\sum_{q=-k}^k |B_{q}^{(k)}|^2 \right]^{1/2}.
\end{equation}

\begin{table}
\caption{Comparison of CFP and the rotational invariants $S_k$ obtained in this paper and CFP
obtained by  Gruber et al. and Burdick et al. (these parameters were rotated in the
coordinate system used in this paper as described in the text).
CFP and $S_k$ are in meV, hybridization parameter equals to 4.76 eV for Nd, 6.8 eV for Sm and Er. }
\begin{tabular}{lrrrrrrrr}
\hline
& & & \multicolumn{2}{c}{Nd}   &\multicolumn{2}{c}{Sm}   &\multicolumn{2}{c}{Er}  \\
& k  & q  & this  paper    & Ref. \onlinecite{burdickNd} &this  paper   & Ref. \onlinecite{gruberSm}  & this  paper   & Ref. \onlinecite{burdickEr} \\
 \hline
& 2 &  0 &    32.8 &  52.3   &    37.0 &    54.3 &    113.1 & 42.3  \\
& 2 &  2 &   -28.7 &  -19.7  &   -23.8 &   -13.3 &     15.4 & -27.6  \\
& 4 &  0 &    -5.8 &   -23.5 &    -1.1 &   -16.1 &   -22.9 &  -21.4 \\
& 4 &  2 &   253.9 &   273.4 &   223.2 &   238.7 &   183.8 & 185.5 \\
& 4 &  4 &  -161.9 &  -119.5 &  -139.5 &   -60.7 &   -87.3 & -52.1  \\
& 6 &  0 &  -228.2 &   -241.1&  -203.9 &  -197.0 &  -207.8 & -146.0 \\
& 6 &  2 &    77.9 &    73.5 &    73.8 &    85.2 &    93.4 &  40.0  \\
& 6 &  4 &   114.9 &   103.5 &   102.4 &   117.2 &    68.3 &  65.6 \\
& 6 &  6 &   108.8 &   72.2 &    95.2 &    85.5 &    54.1 &   64.7 \\
 \hline
    &  2 &   23.3 &   26.5 &   22.4 &   25.7  &  51.5 &   25.7 \\
$S_k$& 4 &  142.0 &  140.9 &  124.1 &  116.2  &  96.2 &   91.1 \\
     & 6 &   93.8 &   88.0 &   83.9 &   85.7  &  76.4 &   56.5 \\
\hline
\end{tabular}
\label{tab:compar}
\end{table}

We now turn to the magnetism, which is more susceptible to the values of CFP compared to the
energies in the zero magnetic field. Despite this sensitivity and the above mentioned problem with  $\Delta$
smaller than 4 eV, the calculated $\hat{g}$ tensors qualitatively reflect the experimental data.
The method will be thus particularly useful for magnetic, nontransparent rare-earth compounds, in which
the experimental data do not allow determination of the crystal field.

Our approach is relatively simple, it may be used by non specialists and corresponding programs, as well as the test example,
are available on the WIEN2k web site (www.wien2k.at).

\section{conclusions}
The crystal field parameters and the $\hat{g}$ tensors of the ground state were calculated for all seven rare-earth Kramers
ions substituted for Y (Lu) in YAG (LuAG). Very good agreement with the spectroscopic data and qualitative agreement with
experimental $\hat{g}$ tensors was found. A comparison of calculated CFP with their counterparts obtained by the
least squares fit to the optical data is fair and it shows that theory can help to avoid the ambiguity inherent to the least
squares approach.  

\acknowledgments
We are indebted to Jan Kune\v{s}, Vladim\'ir Nekvasil and Karel Kn\'i\v{z}ek for helpful discussions.
This work was supported by the Czech Science Foundation 13-09876S project.

\bibliography{garnets}

\end{document}